\documentclass[10pt]{article}
\usepackage{graphicx}
\usepackage{amsmath}
\usepackage{amssymb}
\usepackage{caption2}
\begin{document}
\bibliographystyle{prsty}
\begin{center}
{\large {\bf \sc{  Analysis of  the $X(3872)$, $Z_c(3900)$ and $Z_c(3885)$ as   axial-vector tetraquark states
with  QCD sum rules }}} \\[2mm]
Zhi-Gang  Wang$^{1}$ \footnote{E-mail: zgwang@aliyun.com.  }, Tao Huang$^{2}$ \footnote{Email: huangtao@ihep.ac.cn}     \\
$^{1}$ Department of Physics, North China Electric Power University, Baoding 071003, P. R. China \\
$^{2}$ Institute of High Energy Physics and Theoretical Physics
Center for Science Facilities, Chinese Academy of Sciences, Beijing 100049, P.R. China
\end{center}

\begin{abstract}
In this article, we  distinguish the charge conjunctions of the interpolating currents,
calculate the contributions of the vacuum condensates up to dimension-10  in a consistent way in the operator product expansion,
study the masses and pole residues of the $J^{PC}=1^{+\pm}$ hidden charmed tetraquark states with the QCD sum rules,
and explore the energy scale dependence  in details for the first time. The predictions $M_{X}=3.87^{+0.09}_{-0.09}\,\rm{GeV}$ and
$M_{Z}=3.91^{+0.11}_{-0.09}\,\rm{GeV}$ support assigning    the $X(3872)$ and $Z_c(3900)$ (or $Z_c(3885)$) as the $1^{++}$
and $1^{+-}$ diquark-antidiquark type tetraquark states, respectively.
\end{abstract}



\section{Introduction}

There are many candidates  with the quantum numbers $J^{PC}=0^{++}$ below $2\,\rm{GeV}$, which
cannot be accommodated in one $\bar{q}q$ nonet. The lowest scalar  nonet  mesons $f_0(600)$, $a_0(980)$, $K^*_0(800)$, $f_0(980) $
are usually taken as the tetraquark
states \cite{Close2002}, and have been studied as  the
diquark-antidiquark type tetraquark states with the QCD sum rules \cite{f980-SR}. The QCD sum rules is a powerful theoretical tool  in studying
the ground state hadrons \cite{SVZ79,Reinders85}. For the light tetraquark states, it is difficult to satisfy the two
criteria of the QCD sum rules:

 $\bullet$ Pole dominance at the phenomenological side;

 $\bullet$ Convergence of the operator product expansion \cite{Wang-NPA}. \\
 For the hidden (or doubly) charmed (or bottom) tetraquark states (or molecular states), it is more easy to satisfy the two criteria.

In 2003, the  Belle collaboration   observed   a narrow charmonium-like state $X(3872)$ in the $\pi^+ \pi^- J/\psi$ mass spectrum in the exclusive decay processes  $B^\pm \to K^\pm \pi^+ \pi^- J/\psi$  \cite{X3872-2003}.
   The evidences for the decay modes $X(3872) \to \gamma J/\psi, \, \gamma \psi^{\prime}$  imply the positive charge conjunction $C=+$ \cite{X3872-Jpsi-gamma}.  Angular correlations between final state particles in the $\pi^+ \pi^- J/\psi$ favor the $J^{PC}=1^{++}$ assignment \cite{X3872-JPC}.
  L. Maiani et al  tentatively identify the $X(3872)$  as the $J^{PC} =1^{++}$ tetraquark state  with the symmetric spin distribution
$[cq]_{S=1}[\bar{c}\bar{q}]_{S=0} + [cq]_{S=0}[\bar{c}\bar{q}]_{S=1}$ \cite{Maiani-3872}. For other possible assignments, one can consult the reviews \cite{Swanson2006}.
In Ref.\cite{Narison-3872}, R. D. Matheus et al take the $X(3872) $ as the $J^{PC}=1^{++}$ diquark-antidiquark type  tetraquark state,
and study its mass with the QCD sum rules   by taking the vacuum condensates up  to dimension-8 in the operator product expansion.
 Thereafter   the hidden charmed (or bottom) and doubly  open charmed (or bottom) diquark-antidiquark type tetraquark states have been studied extensively
 with the QCD sum rules \cite{cc-tetraquark,cc-Wang,Wang-Axial,CFQiao1307}. For some articles on the QCD sum rules for the hidden
  charmed (or bottom)  molecular states,
 one can consult the reviews \cite{Heavy quarkonium}.

In 2013, the BESIII collaboration studied  the process  $e^+e^- \to \pi^+\pi^-J/\psi$ at a center-of-mass energy of 4.260 GeV using a $525\, \rm{pb}^{-1}$ data
sample collected with the BESIII detector, and observed a structure $Z_c(3900)$ in the $\pi^\pm J/\psi$ mass spectrum with a mass of $(3899.0\pm 3.6\pm 4.9)\,\rm{ MeV}$
and a width of $(46\pm 10\pm 20) \,\rm{MeV}$ \cite{BES3900}. Then the structure $Z_c(3900)$ was confirmed by the Belle and CLEO collaborations \cite{Belle3900,CLEO3900}.
 R. Faccini et al tentatively identify the $Z_c(3900)$ as the negative charge conjunction partner of the $X(3872)$ \cite{Maiani1303}, other assignments,
 such as molecular state \cite{Molecular3900}, tetraquark state \cite{Tetraquark3900}, hadro-charmonium \cite{hadro-charmonium-3900}, rescattering effect \cite{FSI3900},
 are also suggested. C. F. Qiao and L. Tang  studied  the $J^{PC}=1^{+-}$  hidden charmed tetraquark state with the QCD sum rules by taking the vacuum condensates up
 to dimension-8 in the operator product expansion, and obtained the mass  $M_{Z} = (3912^{+106}_{-103}) \,  \rm{MeV}$ \cite{CFQiao1307}.

Recently, the BESIII collaboration studied  the process $e^+e^- \to \pi D \bar{D}^*$ at $\sqrt{s}=4.26\,\rm{ GeV}$ using a $525\, \rm{pb}^{-1}$ data sample
collected with the BESIII detector at the BEPCII storage ring, and observed a distinct charged structure $Z_c(3885)$  in the $(D \bar{D}^*)^{\pm}$
invariant mass distribution \cite{BES-3885}. The measured mass and width are $(3883.9 \pm 1.5 \pm 4.2)\,\rm{ MeV}$ and  $(24.8 \pm 3.3 \pm 11.0)\,\rm{ MeV}$,
respectively, and   the angular distribution of the $\pi Z_c(3885)$ system favors a $J^P=1^+$ assignment \cite{BES-3885}. We tentatively  identify the $Z_c(3900)$
 and $Z_c(3885)$ as the same particle according to  the uncertainties of the masses and widths. The $1^+$ hidden charmed tetraquark states can decay to
  both the $D\bar{D}^*$ and $\pi J/\psi$ final states.

In the QCD sum rules for the hidden charmed (or bottom) tetraquark states and molecular states, the integrals
 \begin{eqnarray}
 \int_{4m_Q^2}^{s_0} ds \rho_{QCD}(s)\exp\left(-\frac{s}{T^2} \right)\, ,
 \end{eqnarray}
are sensitive to the heavy quark masses $m_Q$, where the $\rho_{QCD}(s)$ denotes the QCD spectral densities and the $T^2$ denotes the Borel parameters.
Variations of the heavy quark masses lead to changes of integral ranges $4m_Q^2-s_0$ of the variable  $\bf{ds}$ besides the QCD spectral densities,
therefore changes of the Borel windows and predicted masses and pole residues. In calculations, we observe that small variations of the heavy quark
masses $m_Q$ can lead to rather large changes of the predictions \cite{cc-Wang,Wang-Axial,Wang4140}.   In previous works, the $\overline{MS}$ masses are taken,
however,  the energy scales at which the QCD spectral densities are calculated are either not shown explicitly
(or not specified) \cite{Narison-3872,cc-tetraquark,CFQiao1307} or shown explicitly at a special value \cite{cc-Wang,Wang-Axial,Wang4140},
the energy scale dependence is not studied in details.

In previous works \cite{Wang-Axial}, we have studied the axial-vector hidden charmed and hidden bottom tetraquark states with the QCD sum rules,
the charge conjunctions are not distinguished. In this article, we  distinguish
the charge conjunctions of the interpolating currents,  calculate the contributions of the vacuum condensates up to dimension-10
in a consistent way in the operator product expansion  and discard the perturbative corrections,
study the masses and pole residues of the axial-vector hidden charmed tetraquark states,
and explore the energy scale dependence in details, and make tentative assignments  of
the $X(3872)$ and $Z_c(3900)$ (or $Z_c(3885)$). In Refs.\cite{Narison-3872,Wang-Axial,CFQiao1307},
some higher dimension vacuum condensates are neglected.
The higher dimension vacuum condensates play an important role in determining the Borel windows,
although they maybe play a less important role in the Borel windows. Different  Borel windows lead to different ground state  masses and pole residues.

 The mass is a fundamental parameter in describing a hadron. In order to identify  the $X(3872)$ and $Z_c(3900)$
 (or $Z_c(3885)$) as the $J^{PC}=1^{++}$ and $1^{+-}$   hidden charmed tetraquark states, respectively,
 we must prove that their  masses lie  in the region $(3.8-4.0)\, \rm{GeV}$ in a consistent way,
 and there exists  a small energy gap between the $C=+$ and $-$ axial-vector tetraquark states.

The article is arranged as follows:  we derive the QCD sum rules for
the masses and pole residues of  the axial-vector tetraquark states  in section 2; in section 3,
we present the numerical results and discussions; section 4 is reserved for our conclusion.

\section{QCD sum rules for  the   $J^{PC}=1^{+\pm}$ tetraquark states }
In the following, we write down  the two-point correlation functions $\Pi_{\mu\nu}(p)$  in the QCD sum rules,
\begin{eqnarray}
\Pi_{\mu\nu}(p)&=&i\int d^4x e^{ip \cdot x} \langle0|T\left\{J_\mu(x)J_\nu^{\dagger}(0)\right\}|0\rangle \, , \\
J_\mu(x)&=&\frac{\epsilon^{ijk}\epsilon^{imn}}{\sqrt{2}}\left\{u^j(x)C\gamma_5c^k(x) \bar{d}^m(x)\gamma_\mu C \bar{c}^n(x)+tu^j(x)C\gamma_\mu c^k(x)\bar{d}^m(x)\gamma_5C \bar{c}^n(x) \right\} \, ,
\end{eqnarray}
$t=\pm 1$ denote the positive and negative charge conjunctions, respectively, the $i$, $j$, $k$, $m$, $n$ are color indexes, the $C$ is the charge conjunction matrix. We choose  the   currents $J_\mu(x)$ to interpolate the
  $J^{PC}=1^{+\pm}$ diquark-antidiquark type tetraquark states $X(3872)$ (to be more precise, the charged partner of the $X(3872)$) and $Z_c(3900)$ (or $Z_c(3885)$), respectively.  Under charge conjunction transform $\widehat{C}$, the currents $J_\mu(x)$ have the properties,
\begin{eqnarray}
\widehat{C}J_{\mu}(x)\widehat{C}^{-1}&=&\pm J_\mu(x) \,\,\,\, {\rm for}\,\,\,\, t=\pm1\, ,
\end{eqnarray}
which originate from the charge conjunction properties of the scalar and axial-vector diquark states,
\begin{eqnarray}
\widehat{C}\left[\epsilon^{ijk}q^j C\gamma_5 c^k\right]\widehat{C}^{-1}&=&\epsilon^{ijk}\bar{q}^j \gamma_5 C \bar{c}^k \, , \nonumber\\
\widehat{C}\left[\epsilon^{ijk}q^j C\gamma_\mu c^k\right]\widehat{C}^{-1}&=&\epsilon^{ijk}\bar{q}^j \gamma_\mu C \bar{c}^k \, .
\end{eqnarray}

We can insert  a complete set of intermediate hadronic states with
the same quantum numbers as the current operators $J_\mu(x)$ into the
correlation functions $\Pi_{\mu\nu}(p)$  to obtain the hadronic representation
\cite{SVZ79,Reinders85}. After isolating the ground state
contributions from the pole terms, which are supposed to be tetraquark states $X(3872)$ and $Z_c(3900)$ (or $Z_c(3885)$), we get the following results,
\begin{eqnarray}
\Pi_{\mu\nu}(p)&=&\frac{\lambda_{X/Z}^2}{M_{X/Z}^2-p^2}\left(-g_{\mu\nu} +\frac{p_\mu p_\nu}{p^2}\right) +\cdots \, \, ,
\end{eqnarray}
where the pole residues (or couplings) $\lambda_{X/Z}$ are defined by
\begin{eqnarray}
 \langle 0|J_\mu(0)|X/Z(p)\rangle=\lambda_{X/Z} \, \varepsilon_\mu \, ,
\end{eqnarray}
the $\varepsilon_\mu$ are the polarization vectors of the axial-vector mesons $X(3872)$ and $Z_c(3900)$ (or $Z_c(3885)$).

The   currents $J_\mu(x)$  have non-vanishing couplings  with the scattering states  $D
D^\ast$, $J/\psi \pi$, $J/\psi \rho$, etc  \cite{PDG}.
 In the following, we
study the contributions of the  intermediate   meson-loops to the correlation functions $\Pi_{\mu\nu}(p)$,
\begin{eqnarray}
\Pi_{\mu\nu}(p)&=&-\frac{\widehat{\lambda}_{X/Z}^{2}}{ p^2-\widehat{M}_{X/Z}^2}\widetilde{g}_{\mu\nu}(p)-\frac{\widehat{\lambda}_{X/Z}}{p^2-\widehat{M}_{X/Z}^2}\widetilde{g}_{\mu\alpha}(p)
 \Sigma_{DD^*}(p) \widetilde{g}^{\alpha\beta}(p) \widetilde{g}_{\beta\nu}(p)\frac{\widehat{\lambda}_{X/Z}}{p^2-\widehat{M}_{X/Z}^2} \nonumber \\
 &&-\frac{\widehat{\lambda}_{X/Z}}{p^2-\widehat{M}_{X/Z}^2}\widetilde{g}_{\mu\alpha}(p)
 \Sigma_{J/\psi\pi}(p) \widetilde{g}^{\alpha\beta}(p) \widetilde{g}_{\beta\nu}(p)\frac{\widehat{\lambda}_{X/Z}}{p^2-\widehat{M}_{X/Z}^2} \nonumber \\
 &&-\frac{\widehat{\lambda}_{X/Z}}{p^2-\widehat{M}_{X/Z}^2}\widetilde{g}_{\mu\alpha}(p)
 \Sigma_{J/\psi\rho}^{\alpha\beta}(p)   \widetilde{g}_{\beta\nu}(p)\frac{\widehat{\lambda}_{X/Z}}{p^2-\widehat{M}_{X/Z}^2} +\cdots \, ,\nonumber \\
 &=&-\frac{\widehat{\lambda}_{X/Z}^{2}}{ p^2-\widehat{M}_{X/Z}^2-\Sigma_{DD^*}(p)-\Sigma_{J/\psi\pi}(p)-\Sigma_{J/\psi\rho}(p)+\cdots}\widetilde{g}_{\mu\nu}(p)+\cdots \, , \end{eqnarray}
where
\begin{eqnarray}
\Sigma_{DD^*}(p)&=&i\int~{d^4q\over(2\pi)^4}\frac{G^2_{X/ZDD^*}}{\left[q^2-M_{D}^2\right]\left[ (p-q)^2-M_{D^*}^2\right]} \, ,\\
\Sigma_{J/\psi \pi}(p)&=&i\int~{d^4q\over(2\pi)^4}\frac{G^2_{X/ZJ/\psi \pi}}{\left[q^2-M_{J/\psi}^2\right]\left[ (p-q)^2-M_{\pi}^2\right]} \, ,\\
\Sigma_{J/\psi\rho}^{\alpha\beta}(p)&=&i\int~{d^4q\over(2\pi)^4}\frac{G^2_{X/ZJ/\psi \rho}\epsilon^{\alpha\theta\sigma\tau}\epsilon^{\beta\theta^{\prime}\sigma^{\prime}\tau^{\prime}}p_\tau p_{\tau^\prime}\widetilde{g}_{\theta\theta^{\prime}}(q)\widetilde{g}_{\sigma\sigma^{\prime}}(p-q)}{\left[
q^2-M_{J/\psi}^2\right]\left[ (p-q)^2-M_{\rho}^2\right]}\, , \nonumber\\
&=&\Sigma_{J/\psi \rho}(p)\widetilde{g}^{\alpha\beta}(p)+\Sigma_{J/\psi \rho}^1(p) \frac{p^{\alpha}p^{\beta} }{p^2}\, ,
\end{eqnarray}
$\widetilde{g}_{\mu\nu}(p)=-g_{\mu\nu}+\frac{p_{\mu}p_{\nu}}{p^2}$, the $G_{X/Z DD^*}$, $G_{X/Z J/\psi\pi}$, $G_{X/Z J/\psi\rho}$ are hadronic coupling constants, the $\widehat{\lambda}_{X/Z}$ and $\widehat{M}_{X/Z}$ are bare quantities to absorb the divergences in the self-energies $\Sigma_{DD^*}(p)$, $\Sigma_{J/\psi \pi}(p)$, $\Sigma_{J/\psi \rho}(p)$, etc.
The renormalized self-energies  contribute  a finite imaginary part to modify the dispersion relation,
\begin{eqnarray}
\Pi_{\mu\nu}(p) &=&-\frac{\lambda_{X/Z}^{2}}{ p^2-M_{X/Z}^2+i\sqrt{p^2}\Gamma(p^2)}\widetilde{g}_{\mu\nu}(p)+\cdots \, ,
 \end{eqnarray}
the physical  widths $\Gamma_{Z_c(3900)}(M_Z^2)=(46 \pm 10 \pm 20)\, \rm{MeV}$ and $\Gamma_{X(3872)}(M_X^2)<1.2\,\rm{MeV}$ are small enough,
 the zero width approximation in  the hadronic spectral densities works.

The contaminations of the intermediate meson-loops are expected
 to be small,  we take a single pole approximation  and approximate the continuum contributions as
\begin{eqnarray}
\int_{4m_c^2}^{\infty} ds \, \frac{1}{s-p^2}\rho_{QCD}(s)\Theta(s-s_0)\, ,
\end{eqnarray}
the $\rho_{QCD}(s)$ denotes the full QCD spectral densities; the pole term embodies  the net
effects.  Onset of the continuum states
is not abrupt,   the ground state, the first excited state, the
second excited state, etc,  the continuum states appear sequentially.
The threshold parameter $s_0$  is postponed to large value, where the spectral densities can be well approximated by the
contributions of the asymptotic quarks and gluons, in other words, the perturbative contributions. If only the ground state is taken, the $s_0$ is not large enough to warrant that the hadronic spectral densities above the $s_0$ can be approximated by the perturbative contributions, the $\rho_{QCD}(s)$ should include the contributions of the vacuum condensates besides the perturbative terms.

 In the following,  we briefly outline  the operator product expansion for the correlation functions $\Pi_{\mu\nu}(p)$  in perturbative
QCD \footnote{It is convenient  to introduce the external fields $\bar{\chi}$, $\chi$, $A_\mu^a$ and additional Lagrangian $\Delta \mathcal{L}$
\begin{eqnarray}
\Delta {\mathcal{L}}&=&\bar{q}(x)i\gamma^\mu \partial_\mu\chi(x)+\bar{\chi}(x)i\gamma^\mu \partial_\mu q(x)+g_s \bar{q}(x)\gamma^\mu t^a q(x)A_\mu^a(x)+\cdots\, , \nonumber
\end{eqnarray}
in carrying out the operator product expansion \cite{Reinders85,external}. We expand the heavy and light quark propagators $S^Q_{ij}$ and $S_{ij}$ in terms of the
external fields $\bar{\chi}$, $\chi$ and $A_\mu^a$,
\begin{eqnarray}
S^Q_{ij}\left(x,\bar{\chi},\chi,A_\mu^a\right)&=&\frac{i}{(2\pi)^4}\int d^4k e^{-ik \cdot x} \left\{
\frac{\delta_{ij}}{\!\not\!{k}-m_Q}
-\frac{g_s A^a_{\alpha\beta}t^a_{ij}}{4}\frac{\sigma^{\alpha\beta}(\!\not\!{k}+m_Q)+(\!\not\!{k}+m_Q)
\sigma^{\alpha\beta}}{(k^2-m_Q^2)^2}+\cdots\right\} \, , \nonumber
\end{eqnarray}
\begin{eqnarray}
S_{ij}\left(x,\bar{\chi},\chi,A_\mu^a\right)&=& \frac{i\delta_{ij}\!\not\!{x}}{ 2\pi^2x^4}+\chi^i(x)\bar{\chi}^j(0) -\frac{ig_s A^{a}_{\alpha\beta}t^a_{ij}\left(\!\not\!{x}
\sigma^{\alpha\beta}+\sigma^{\alpha\beta} \!\not\!{x}\right)}{32\pi^2x^2}+\cdots \, ,\nonumber
\end{eqnarray}
where $A^a_{\alpha\beta}=\partial_\alpha A^a_\beta-\partial_\beta A^a_\alpha+g_s f^{abc}A_\alpha^b A_\beta^c$.
Then the  correlation functions $\Pi(p)$ can be written as
\begin{eqnarray}
\Pi(p)&=& \sum_{n=0}^\infty {\mathcal{C}}_n(p)\,\, {\mathcal{O}}_n\left(\bar{\chi},\chi,A_\mu^a\right)\, , \nonumber
\end{eqnarray}
in the external fields $\bar{\chi}$, $\chi$ and $A_\mu^a$, where the ${\mathcal{C}}_n(p)$ are the Wilson's coefficients,   the operators  ${\mathcal{O}}_n\left(\bar{\chi},\chi,A_\mu^a\right)$ are characterized by their dimensions $n$. If we neglect the perturbative corrections, the
 operators  ${\mathcal{O}}_n\left(\bar{\chi},\chi,A_\mu^a\right)$ can also be counted by the orders of the fine structure  constant $\alpha_s=\frac{g_s^2}{4\pi}$, ${\mathcal{O}}\left(\alpha_s^k\right)$,
  with $k=0, \frac{1}{2}, 1, \frac{3}{2}$, etc. In this article, we take the truncations $n\leq 10$ and $k\leq 1$, and
  factorize the higher dimension operators into non-factorizable
   low dimension operators with the same quantum numbers of the vacuum.
  Taking the following replacements
  \begin{eqnarray}
  {\mathcal{O}}_n\left(\bar{\chi},\chi,A_\mu^a\right) &\to& \langle{\mathcal{O}}_n\left(\bar{q},q,G_\mu^a\right)\rangle \, ,\nonumber
  \end{eqnarray}
  we obtain the correlation functions at the level of quark-gluon degrees of freedom. For example,
  \begin{eqnarray}
    \chi^i(x)\bar{\chi}^j(0)  &=& -\frac{\delta_{ij}
\bar{\chi}(0)\chi(0) }{12} -\frac{\delta_{ij}x^2  \bar{\chi}(0)g_s\sigma A(0)\chi(0) }{192}+\cdots\to -\frac{\delta_{ij}\langle
\bar{q}q\rangle}{12} -\frac{\delta_{ij}x^2\langle \bar{q}g_s\sigma Gq\rangle}{192}+\cdots\, . \nonumber
  \end{eqnarray}
   For simplicity, we often  take the following replacements,
  \begin{eqnarray}
  S^Q_{ij}\left(x,\bar{\chi},\chi,A_\mu^a\right)&\to&S^Q_{ij}\left(x,\bar{q},q,G_\mu^a\right)\, , \nonumber\\
  S_{ij}\left(x,\bar{\chi},\chi,A_\mu^a\right)&\to&S_{ij}\left(x,\bar{q},q,G_\mu^a\right)\, , \nonumber\\
 {\mathcal{O}}_n\left(\bar{\chi},\chi,A_\mu^a\right) &\to& \langle {\mathcal{O}}_n\left(\bar{q},q,G_\mu^a\right)\rangle\, , \nonumber
  \end{eqnarray}
directly in calculations by neglecting some intermediate steps, and resort to the routine  taken in this article.}.  We contract the quark fields in the correlation functions
$\Pi_{\mu\nu}(p)$ with Wick theorem, obtain the results:
\begin{eqnarray}
\Pi_{\mu\nu}(p)&=&-\frac{i\epsilon^{ijk}\epsilon^{imn}\epsilon^{i^{\prime}j^{\prime}k^{\prime}}\epsilon^{i^{\prime}m^{\prime}n^{\prime}}}{2}\int d^4x e^{ip \cdot x}   \nonumber\\
&&\left\{{\rm Tr}\left[ \gamma_5C^{kk^{\prime}}(x)\gamma_5 CU^{jj^{\prime}T}(x)C\right] {\rm Tr}\left[ \gamma_\nu C^{n^{\prime}n}(-x)\gamma_\mu C D^{m^{\prime}mT}(-x)C\right] \right. \nonumber\\
&&+{\rm Tr}\left[ \gamma_\mu C^{kk^{\prime}}(x)\gamma_\nu CU^{jj^{\prime}T}(x)C\right] {\rm Tr}\left[ \gamma_5 C^{n^{\prime}n}(-x)\gamma_5 C D^{m^{\prime}mT}(-x)C\right] \nonumber\\
&&\mp{\rm Tr}\left[ \gamma_\mu C^{kk^{\prime}}(x)\gamma_5 CU^{jj^{\prime}T}(x)C\right] {\rm Tr}\left[ \gamma_\nu C^{n^{\prime}n}(-x)\gamma_5 C D^{m^{\prime}mT}(-x)C\right] \nonumber\\
 &&\left.\mp{\rm Tr}\left[ \gamma_5 C^{kk^{\prime}}(x)\gamma_\nu CU^{jj^{\prime}T}(x)C\right] {\rm Tr}\left[ \gamma_5 C^{n^{\prime}n}(-x)\gamma_\mu C D^{m^{\prime}mT}(-x)C\right] \right\} \, ,
\end{eqnarray}
where the $\mp$ correspond  the positive and negative charge conjunctions, respectively,
 the $U_{ij}(x)$, $D_{ij}(x)$ and $C_{ij}(x)$ are the full $u$, $d$ and $c$ quark propagators, respectively (the $U_{ij}(x)$ and $D_{ij}(x)$ can be written as $S_{ij}(x)$ for simplicity),
\begin{eqnarray}
S_{ij}(x)&=& \frac{i\delta_{ij}\!\not\!{x}}{ 2\pi^2x^4}-\frac{\delta_{ij}\langle
\bar{q}q\rangle}{12} -\frac{\delta_{ij}x^2\langle \bar{q}g_s\sigma Gq\rangle}{192} -\frac{ig_sG^{a}_{\alpha\beta}t^a_{ij}\left(\!\not\!{x}
\sigma^{\alpha\beta}+\sigma^{\alpha\beta} \!\not\!{x}\right)}{32\pi^2x^2} -\frac{i\delta_{ij}x^2\!\not\!{x}g_s^2\langle \bar{q} q\rangle^2}{7776}\nonumber\\
&&  -\frac{\delta_{ij}x^4\langle \bar{q}q \rangle\langle g_s^2 GG\rangle}{27648} -\frac{1}{8}\langle\bar{q}_j\sigma^{\mu\nu}q_i \rangle \sigma_{\mu\nu}-\frac{1}{4}\langle\bar{q}_j\gamma^{\mu}q_i\rangle \gamma_{\mu }+\cdots \, ,
\end{eqnarray}
\begin{eqnarray}
C_{ij}(x)&=&\frac{i}{(2\pi)^4}\int d^4k e^{-ik \cdot x} \left\{
\frac{\delta_{ij}}{\!\not\!{k}-m_c}
-\frac{g_sG^n_{\alpha\beta}t^n_{ij}}{4}\frac{\sigma^{\alpha\beta}(\!\not\!{k}+m_c)+(\!\not\!{k}+m_c)
\sigma^{\alpha\beta}}{(k^2-m_c^2)^2}\right.\nonumber\\
&&\left. +\frac{g_s D_\alpha G^n_{\beta\lambda}t^n_{ij}(f^{\lambda\beta\alpha}+f^{\lambda\alpha\beta}) }{3(k^2-m_c^2)^4}
-\frac{g_s^2 (t^at^b)_{ij} G^a_{\alpha\beta}G^b_{\mu\nu}(f^{\alpha\beta\mu\nu}+f^{\alpha\mu\beta\nu}+f^{\alpha\mu\nu\beta}) }{4(k^2-m_c^2)^5}+\cdots\right\} \, , \nonumber \\
\end{eqnarray}
\begin{eqnarray}
f^{\lambda\alpha\beta}&=&(\!\not\!{k}+m_c)\gamma^\lambda(\!\not\!{k}+m_c)\gamma^\alpha(\!\not\!{k}+m_c)\gamma^\beta(\!\not\!{k}+m_c)\, ,\nonumber\\
f^{\alpha\beta\mu\nu}&=&(\!\not\!{k}+m_c)\gamma^\alpha(\!\not\!{k}+m_c)\gamma^\beta(\!\not\!{k}+m_c)\gamma^\mu(\!\not\!{k}+m_c)\gamma^\nu(\!\not\!{k}+m_c)\, ,
\end{eqnarray}
and  $t^n=\frac{\lambda^n}{2}$, the $\lambda^n$ is the Gell-Mann matrix,  $D_\alpha=\partial_\alpha-ig_sG^n_\alpha t^n$ \cite{Reinders85},
then compute  the integrals both in the coordinate and momentum spaces,  and obtain the correlation functions $\Pi_{\mu\nu}(p)$ therefore the spectral densities at the level of   quark-gluon degrees  of freedom.  The condensates $g_s^2\langle \bar{q} q\rangle^2$ and $\langle \bar{q}q \rangle\langle g_s^2 GG\rangle$ in the full light-quark propagators $S_{ij}(x)$ come from the Taylor expansion in terms of the covariant derivatives,
\begin{eqnarray}
q(x)&=&\sum_{n=0}^{\infty} \frac{1}{n!}\,x^{\mu_1}x^{\mu_2}\cdots x^{\mu_n}\, D_{\mu_1}D_{\mu_2} \cdots D_{\mu_n}\, q(0)\, , \nonumber\\
\bar{q}(x)&=&\sum_{n=0}^{\infty} \frac{1}{n!}\,x^{\mu_1}x^{\mu_2}\cdots x^{\mu_n}\, \bar{q}(0)\, D^{\dagger}_{\mu_1}D^{\dagger}_{\mu_2} \cdots D^{\dagger}_{\mu_n} \, ,
\end{eqnarray}
with $n=3$ and $n=4$, respectively.
In Eq.(15), we retain the terms $\langle\bar{q}_j\sigma_{\mu\nu}q_i \rangle$ and $\langle\bar{q}_j\gamma_{\mu}q_i\rangle$ originate from the Fierz re-ordering of the $\langle q_i \bar{q}_j\rangle$ to  absorb the gluons  emitted from the heavy quark lines to form $\langle\bar{q}_j g_s G^a_{\alpha\beta} t^a_{mn}\sigma_{\mu\nu} q_i \rangle$ and $\langle\bar{q}_j\gamma_{\mu}q_ig_s D_\nu G^a_{\alpha\beta}t^a_{mn}\rangle$ so as to extract the mixed condensate and four-quark condensates $\langle\bar{q}g_s\sigma G q\rangle$ and $g_s^2\langle\bar{q}q\rangle^2$, respectively, see the typical Feynman diagrams shown in Fig.1.

\begin{figure}
 \centering
 \includegraphics[totalheight=4cm,width=14cm]{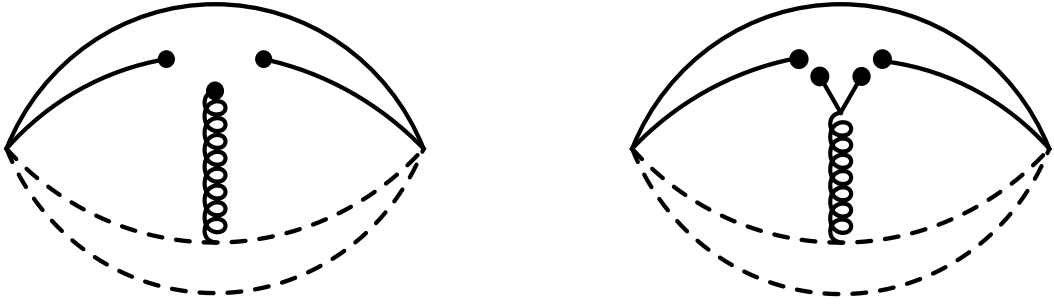}
    \caption{The typical Feynman diagrams contribute to the mixed condensates and four-quark condensates, where the solid and dashed lines denote the light and heavy quark propagators, respectively. }
\end{figure}

 Once analytical results are obtained,  we can take the
quark-hadron duality and perform Borel transform  with respect to
the variable $P^2=-p^2$ to obtain  the following QCD sum rules:
\begin{eqnarray}
\lambda_{X/Z}^2\, \exp\left(-\frac{M_{X/Z}^2}{T^2}\right)= \int_{4m_c^2}^{s_0} ds\, \rho(s) \, \exp\left(-\frac{s}{T^2}\right) \, ,
\end{eqnarray}
where
\begin{eqnarray}
\rho(s)&=&\rho_{0}(s)+\rho_{3}(s) +\rho_{4}(s)+\rho_{5}(s)+\rho_{6}(s)+\rho_{7}(s) +\rho_{8}(s)+\rho_{10}(s)\, ,
\end{eqnarray}

\begin{eqnarray}
\rho_{0}(s)&=&\frac{1}{3072\pi^6}\int_{y_i}^{y_f}dy \int_{z_i}^{1-y}dz \, yz(1-y-z)^3\left(s-\overline{m}_c^2\right)^2\left(35s^2-26s\overline{m}_c^2+3\overline{m}_c^2 \right)\, , \nonumber
\end{eqnarray}

\begin{eqnarray}
\rho_{3}(s)&=&-\frac{m_c\langle \bar{q}q\rangle}{64\pi^4}\int_{y_i}^{y_f}dy \int_{z_i}^{1-y}dz \, (y+z)(1-y-z)\left(s-\overline{m}_c^2\right)\left(7s-3\overline{m}_c^2 \right) \, , \nonumber
\end{eqnarray}

\begin{eqnarray}
\rho_{4}(s)&=&-\frac{m_c^2}{2304\pi^4} \langle\frac{\alpha_s GG}{\pi}\rangle\int_{y_i}^{y_f}dy \int_{z_i}^{1-y}dz \left( \frac{z}{y^2}+\frac{y}{z^2}\right)(1-y-z)^3 \left\{ 8s-3\overline{m}_c^2+\overline{m}_c^4\delta\left(s-\overline{m}_c^2\right)\right\} \nonumber\\
&&+\frac{1}{1536\pi^4}\langle\frac{\alpha_s GG}{\pi}\rangle\int_{y_i}^{y_f}dy \int_{z_i}^{1-y}dz (y+z)(1-y-z)^2 \,s\,(5s-4\overline{m}_c^2) \nonumber\\
&&-t\frac{m_c^2}{1152\pi^4}\langle\frac{\alpha_s GG}{\pi}\rangle\int_{y_i}^{y_f}dy \int_{z_i}^{1-y}dz \left(s-\overline{m}_c^2 \right)\left\{ 1-\left( \frac{1}{y}+ \frac{1}{z}\right) (1-y-z) \right. \nonumber\\
&&\left.+ \frac{(1-y-z)^2}{2yz}  -\frac{1-y-z}{2} +\left(\frac{1}{y}+\frac{1}{z} \right)\frac{(1-y-z)^2}{4}
 -\frac{(1-y-z)^3}{12yz}  \right\}\, ,\nonumber
\end{eqnarray}

\begin{eqnarray}
\rho_{5}(s)&=&\frac{m_c\langle \bar{q}g_s\sigma Gq\rangle}{128\pi^4}\int_{y_i}^{y_f}dy \int_{z_i}^{1-y}dz   (y+z) \left(5s-3\overline{m}_c^2 \right) \nonumber\\
&&-\frac{m_c\langle \bar{q}g_s\sigma Gq\rangle}{128\pi^4}\int_{y_i}^{y_f}dy \int_{z_i}^{1-y}dz   \left(\frac{y}{z}+\frac{z}{y} \right)(1-y-z) \left(2s-\overline{m}_c^2 \right)  \nonumber\\
&&-t\frac{m_c\langle \bar{q}g_s\sigma Gq\rangle}{1152\pi^4}\int_{y_i}^{y_f}dy \int_{z_i}^{1-y}dz   \left(\frac{y}{z}+\frac{z}{y} \right)(1-y-z) \left(5s-3\overline{m}_c^2 \right) \, , \nonumber
\end{eqnarray}

\begin{eqnarray}
\rho_{6}(s)&=&\frac{m_c^2\langle\bar{q}q\rangle^2}{12\pi^2}\int_{y_i}^{y_f}dy +\frac{g_s^2\langle\bar{q}q\rangle^2}{648\pi^4}\int_{y_i}^{y_f}dy \int_{z_i}^{1-y}dz\, yz \left\{8s-3\overline{m}_c^2 +\overline{m}_c^4\delta\left(s-\overline{m}_c^2 \right)\right\}\nonumber\\
&&-\frac{g_s^2\langle\bar{q}q\rangle^2}{2592\pi^4}\int_{y_i}^{y_f}dy \int_{z_i}^{1-y}dz(1-y-z)\left\{ \left(\frac{z}{y}+\frac{y}{z} \right)3\left(7s-4\overline{m}_c^2 \right)\right.\nonumber\\
&&\left.+\left(\frac{z}{y^2}+\frac{y}{z^2} \right)m_c^2\left[ 7+5\overline{m}_c^2\delta\left(s-\overline{m}_c^2 \right)\right]-(y+z)\left(4s-3\overline{m}_c^2 \right)\right\} \nonumber\\
&&-\frac{g_s^2\langle\bar{q}q\rangle^2}{3888\pi^4}\int_{y_i}^{y_f}dy \int_{z_i}^{1-y}dz(1-y-z)\left\{  \left(\frac{z}{y}+\frac{y}{z} \right)3\left(2s-\overline{m}_c^2 \right)\right. \nonumber\\
&&\left.+\left(\frac{z}{y^2}+\frac{y}{z^2} \right)m_c^2\left[ 1+\overline{m}_c^2\delta\left(s-\overline{m}_c^2\right)\right]+(y+z)2\left[8s-3\overline{m}_c^2 +\overline{m}_c^4\delta\left(s-\overline{m}_c^2\right)\right]\right\}\, ,\nonumber
\end{eqnarray}

\begin{eqnarray}
\rho_7(s)&=&\frac{m_c^3\langle\bar{q}q\rangle}{576\pi^2}\langle\frac{\alpha_sGG}{\pi}\rangle\int_{y_i}^{y_f}dy \int_{z_i}^{1-y}dz \left(\frac{y}{z^3}+\frac{z}{y^3}+\frac{1}{y^2}+\frac{1}{z^2}\right)(1-y-z) \nonumber\\
&&\left( 1+\frac{2\overline{m}_c^2}{T^2}\right)\delta\left(s-\overline{m}_c^2\right)\nonumber\\
&&-\frac{m_c\langle\bar{q}q\rangle}{64\pi^2}\langle\frac{\alpha_sGG}{\pi}\rangle\int_{y_i}^{y_f}dy \int_{z_i}^{1-y}dz \left(\frac{y}{z^2}+\frac{z}{y^2}\right)(1-y-z)
\left\{1+\frac{2\overline{m}_c^2}{3}\delta\left(s-\overline{m}_c^2\right) \right\} \nonumber\\
&&-\frac{m_c\langle\bar{q}q\rangle}{192\pi^2}\langle\frac{\alpha_sGG}{\pi}\rangle\int_{y_i}^{y_f}dy \int_{z_i}^{1-y}dz\left\{1+\frac{2\overline{m}_c^2}{3}\delta\left(s-\overline{m}_c^2\right) \right\} \nonumber\\
&&-t\frac{m_c\langle\bar{q}q\rangle}{288\pi^2}\langle\frac{\alpha_sGG}{\pi}\rangle\int_{y_i}^{y_f}dy \int_{z_i}^{1-y}dz\left\{1-\left(\frac{1}{y}+\frac{1}{z}\right)\frac{1-y-z}{2}\right\}\left\{1+\frac{2\overline{m}_c^2}{3}\delta\left(s-\overline{m}_c^2\right) \right\} \nonumber\\
&&-\frac{m_c\langle\bar{q}q\rangle}{384\pi^2}\langle\frac{\alpha_sGG}{\pi}\rangle\int_{y_i}^{y_f}dy \left\{1+\frac{2\widetilde{m}_c^2}{3}\delta\left(s-\widetilde{m}_c^2\right) \right\}  \, , \nonumber
\end{eqnarray}

\begin{eqnarray}
\rho_8(s)&=&-\frac{m_c^2\langle\bar{q}q\rangle\langle\bar{q}g_s\sigma Gq\rangle}{24\pi^2}\int_0^1 dy \left(1+\frac{\widetilde{m}_c^2}{T^2} \right)\delta\left(s-\widetilde{m}_c^2\right)\nonumber\\
&&+\frac{m_c^2\langle\bar{q}q\rangle\langle\bar{q}g_s\sigma Gq\rangle}{96\pi^2}\int_0^1 dy \left( \frac{1}{y}+\frac{1}{1-y} \right)\delta\left(s-\widetilde{m}_c^2\right)\nonumber\\
&&+t\frac{\langle\bar{q}q\rangle\langle\bar{q}g_s\sigma Gq\rangle}{288\pi^2}\int_{y_i}^{y_f} dy \left\{1+\frac{2\widetilde{m}_c^2}{3}\delta\left(s-\widetilde{m}_c^2\right) \right\}  \, , \nonumber
\end{eqnarray}

\begin{eqnarray}
\rho_{10}(s)&=&\frac{m_c^2\langle\bar{q}g_s\sigma Gq\rangle^2}{192\pi^2T^6}\int_0^1 dy \widetilde{m}_c^4\delta \left( s-\widetilde{m}_c^2\right)\nonumber\\
&&-\frac{m_c^4\langle\bar{q}q\rangle^2}{216T^4}\langle\frac{\alpha_sGG}{\pi}\rangle\int_0^1 dy  \left\{ \frac{1}{y^3}+\frac{1}{(1-y)^3}\right\} \delta\left( s-\widetilde{m}_c^2\right)\nonumber\\
&&+\frac{m_c^2\langle\bar{q}q\rangle^2}{72T^2}\langle\frac{\alpha_sGG}{\pi}\rangle\int_0^1 dy  \left\{ \frac{1}{y^2}+\frac{1}{(1-y)^2}\right\} \delta\left( s-\widetilde{m}_c^2\right)\nonumber\\
&&-t\frac{\langle\bar{q}q\rangle^2}{1296}\langle\frac{\alpha_sGG}{\pi}\rangle\int_0^1 dy  \left( 1+\frac{2\widetilde{m}_c^2}{T^2}\right) \delta\left( s-\widetilde{m}_c^2\right)\nonumber\\
&&-\frac{m_c^2\langle\bar{q}g_s\sigma Gq\rangle^2}{384\pi^2T^4}\int_0^1 dy \left( \frac{1}{y}+\frac{1}{1-y}\right)\widetilde{m}_c^2\delta \left( s-\widetilde{m}_c^2\right)\nonumber\\
&&-t\frac{\langle\bar{q}g_s\sigma Gq\rangle^2}{1728\pi^2}\int_0^1 dy \left(1+\frac{3\widetilde{m}_c^2}{2T^2}+\frac{\widetilde{m}_c^4}{T^4} \right)\delta \left( s-\widetilde{m}_c^2\right)\nonumber\\
&&-t\frac{\langle\bar{q}g_s\sigma Gq\rangle^2}{2304\pi^2}\int_0^1 dy \left(1+\frac{2\widetilde{m}_c^2}{T^2}  \right)\delta \left( s-\widetilde{m}_c^2\right)\nonumber\\
&&+\frac{m_c^2\langle\bar{q}q\rangle^2}{216T^6}\langle\frac{\alpha_sGG}{\pi}\rangle\int_0^1 dy  \widetilde{m}_c^4  \delta\left( s-\widetilde{m}_c^2\right) \, ,
\end{eqnarray}
the subscripts  $0$, $3$, $4$, $5$, $6$, $7$, $8$, $10$ denote the dimensions of the  vacuum condensates, $y_{f}=\frac{1+\sqrt{1-4m_c^2/s}}{2}$,
$y_{i}=\frac{1-\sqrt{1-4m_c^2/s}}{2}$, $z_{i}=\frac{ym_c^2}{y s -m_c^2}$, $\overline{m}_c^2=\frac{(y+z)m_c^2}{yz}$,
$ \widetilde{m}_c^2=\frac{m_c^2}{y(1-y)}$, $\int_{y_i}^{y_f}dy \to \int_{0}^{1}dy$, $\int_{z_i}^{1-y}dz \to \int_{0}^{1-y}dz$ when the $\delta$ functions $\delta\left(s-\overline{m}_c^2\right)$ and $\delta\left(s-\widetilde{m}_c^2\right)$ appear. In calculating the Feynman diagrams, we encounter the terms containing
$\langle \bar{q}\gamma_\mu t^a q g_s D_\eta G^a_{\lambda\tau}\rangle$, $\langle  \bar{q}^{j^{\prime}}g_sG^a_{\alpha\beta}t^a_{kk^{\prime}}\sigma_{\lambda\tau}q^j\rangle$,
$\langle \bar{q}^{m}g_sG^a_{\alpha\beta}t^a_{kk^{\prime}}\sigma_{\lambda\tau}q^{m^{\prime}}\rangle$ and deal them with the following tricks:
\begin{eqnarray}
\langle \bar{q}\gamma_\mu t^a q g_s D_\eta G^a_{\lambda\tau}\rangle &=& \frac{g_{\eta\lambda}g_{\tau\mu}-g_{\eta\tau}g_{\lambda\mu}}{12}\langle \bar{q}\gamma_\rho t^a q g_s D_\sigma G^a_{\sigma\rho}\rangle\, ,\nonumber\\
&=&-\frac{g_{\eta\lambda}g_{\tau\mu}-g_{\eta\tau}g_{\lambda\mu}}{12}g_s^2\langle \bar{q}\gamma_\rho t^a q \sum {}_{\psi=u,d,s}\bar{\psi}\gamma^\rho t^a \psi \rangle \, , \nonumber\\
&=&\frac{g_{\eta\lambda}g_{\tau\mu}-g_{\eta\tau}g_{\lambda\mu}}{27}g_s^2\langle \bar{q}q\rangle^2\, ,
\end{eqnarray}
according to the equation of motion $D^{\nu}G_{\mu\nu}^a=\sum_{\psi=u,d,s}g_s\bar{\psi}\gamma_{\mu}t^a \psi $, and
\begin{eqnarray}
\langle  \bar{q}^{j^{\prime}}g_sG^a_{\alpha\beta}t^a_{kk^{\prime}}\sigma_{\lambda\tau}q^j\rangle\epsilon^{ijk}\epsilon^{i^{\prime}j^{\prime}k^{\prime}}\epsilon^{imn}\epsilon^{i^{\prime}m^{\prime}n^{\prime}}
&=&\frac{\langle \bar{q} g_sG_{\alpha\beta} \sigma_{\lambda\tau}q\rangle}{6}\epsilon^{ijk}\epsilon^{i^{\prime}j^{\prime}k^{\prime}} \epsilon^{imn}\epsilon^{i^{\prime}m^{\prime}n^{\prime}} \delta^{kj^{\prime}}\delta^{jk^{\prime}} \, , \nonumber\\
\langle \bar{q}^{m}g_sG^a_{\alpha\beta}t^a_{kk^{\prime}}\sigma_{\lambda\tau}q^{m^{\prime}}\rangle\epsilon^{ijk}\epsilon^{i^{\prime}j^{\prime}k^{\prime}}\epsilon^{imn}\epsilon^{i^{\prime}m^{\prime}n^{\prime}}
&=&\frac{\langle \bar{q} g_sG_{\alpha\beta} \sigma_{\lambda\tau}q\rangle}{9}\epsilon^{ijk}\epsilon^{i^{\prime}j^{\prime}k^{\prime}} \epsilon^{imn}\epsilon^{i^{\prime}m^{\prime}n^{\prime}} \delta^{km}\delta^{k^{\prime}m^{\prime}} \, , \nonumber\\
\end{eqnarray}
according to antisymmetry property of the three colors.

 In this article, we carry out the
operator product expansion to the vacuum condensates adding up to dimension-10 and discard the  perturbative corrections, and
take the assumption of vacuum saturation for the  higher dimension vacuum condensates.
The condensates $\langle \frac{\alpha_s}{\pi}GG\rangle$, $\langle \bar{q}q\rangle\langle \frac{\alpha_s}{\pi}GG\rangle$,
$\langle \bar{q}q\rangle^2\langle \frac{\alpha_s}{\pi}GG\rangle$, $\langle \bar{q} g_s \sigma Gq\rangle^2$ and $g_s^2\langle \bar{q}q\rangle^2$ are the vacuum expectations
of the operators of the order
$\mathcal{O}(\alpha_s)$.  The four-quark condensate $g_s^2\langle \bar{q}q\rangle^2$ comes from the terms
$\langle \bar{q}\gamma_\mu t^a q g_s D_\eta G^a_{\lambda\tau}\rangle$, $\langle\bar{q}_jD^{\dagger}_{\mu}D^{\dagger}_{\nu}D^{\dagger}_{\alpha}q_i\rangle$  and
$\langle\bar{q}_jD_{\mu}D_{\nu}D_{\alpha}q_i\rangle$, rather than comes from the perturbative corrections of $\langle \bar{q}q\rangle^2$.
 The condensates $\langle g_s^3 GGG\rangle$, $\langle \frac{\alpha_s GG}{\pi}\rangle^2$,
 $\langle \frac{\alpha_s GG}{\pi}\rangle\langle \bar{q} g_s \sigma Gq\rangle$ have the dimensions 6, 8, 9 respectively,  but they are   the vacuum expectations
of the operators of the order    $\mathcal{O}( \alpha_s^{3/2})$, $\mathcal{O}(\alpha_s^2)$, $\mathcal{O}( \alpha_s^{3/2})$ respectively, and discarded.  We take
the truncations $n\leq 10$ and $k\leq 1$ in a consistent way,
the operators of the orders $\mathcal{O}( \alpha_s^{k})$ with $k> 1$ are  discarded. Furthermore,  the values of the  condensates $\langle g_s^3 GGG\rangle$, $\langle \frac{\alpha_s GG}{\pi}\rangle^2$,
 $\langle \frac{\alpha_s GG}{\pi}\rangle\langle \bar{q} g_s \sigma Gq\rangle$   are very small, and they can be  neglected safely.

 Differentiate   Eq.(19) with respect to  $\frac{1}{T^2}$, then eliminate the
 pole residues $\lambda_{X/Z}$, we obtain the QCD sum rules for
 the masses of the $X(3872)$ and $Z_c(3900)$ (or $Z_c(3885)$),
 \begin{eqnarray}
 M^2_{X/Z}= \frac{\int_{4m_c^2}^{s_0} ds\frac{d}{d \left(-1/T^2\right)}\,\rho(s)\exp\left(-\frac{s}{T^2}\right)}{\int_{4m_c^2}^{s_0} ds \,\rho(s)
 \exp\left(-\frac{s}{T^2}\right)}\, .
\end{eqnarray}

\section{Numerical results and discussions}
The input parameters are taken to be the standard values $\langle
\bar{q}q \rangle=-(0.24\pm 0.01\, \rm{GeV})^3$,   $\langle
\bar{q}g_s\sigma G q \rangle=m_0^2\langle \bar{q}q \rangle$,
$m_0^2=(0.8 \pm 0.1)\,\rm{GeV}^2$, $\langle \frac{\alpha_s
GG}{\pi}\rangle=(0.33\,\rm{GeV})^4 $    at the energy scale  $\mu=1\, \rm{GeV}$
\cite{SVZ79,Reinders85,Ioffe2005,ColangeloReview}.
The quark condensate and mixed quark condensate evolve with the   renormalization group equation, $\langle\bar{q}q \rangle(\mu^2)=\langle\bar{q}q \rangle(Q^2)\left[\frac{\alpha_{s}(Q)}{\alpha_{s}(\mu)}\right]^{\frac{4}{9}}$ and $\langle\bar{q}g_s \sigma Gq \rangle(\mu^2)=\langle\bar{q}g_s \sigma Gq \rangle(Q^2)\left[\frac{\alpha_{s}(Q)}{\alpha_{s}(\mu)}\right]^{\frac{2}{27}}$.

In the article, we take the $\overline{MS}$ mass $m_{c}(m_c^2)=(1.275\pm0.025)\,\rm{GeV}$
 from the Particle Data Group \cite{PDG}, and take into account
the energy-scale dependence of  the $\overline{MS}$ mass from the renormalization group equation,
\begin{eqnarray}
m_c(\mu^2)&=&m_c(m_c^2)\left[\frac{\alpha_{s}(\mu)}{\alpha_{s}(m_c)}\right]^{\frac{12}{25}} \, ,\nonumber\\
\alpha_s(\mu)&=&\frac{1}{b_0t}\left[1-\frac{b_1}{b_0^2}\frac{\log t}{t} +\frac{b_1^2(\log^2{t}-\log{t}-1)+b_0b_2}{b_0^4t^2}\right]\, ,
\end{eqnarray}
  where $t=\log \frac{\mu^2}{\Lambda^2}$, $b_0=\frac{33-2n_f}{12\pi}$, $b_1=\frac{153-19n_f}{24\pi^2}$, $b_2=\frac{2857-\frac{5033}{9}n_f+\frac{325}{27}n_f^2}{128\pi^3}$,  $\Lambda=213\,\rm{MeV}$, $296\,\rm{MeV}$  and  $339\,\rm{MeV}$ for the flavors  $n_f=5$, $4$ and $3$, respectively  \cite{PDG}.

Now, we take a short digression to discuss the energy scale dependence of the $c\bar{q}$ and $c\bar{c}$ systems,   and
 write down the QCD sum rules for the $D$  and $J/\psi$ mesons,
 \begin{eqnarray}
 \frac{f_{D}^2 M_{D}^4}{m_c^2}\exp\left(-\frac{M_{D}^2}{T^2}\right)&=& \frac{3}{8\pi^2} \int_{m_c^2}^{s_0} ds s\left(1-\frac{m_c^2}{s}\right)^2  \exp\left(-\frac{s}{T^2}\right) -m_c\langle\bar{q}q\rangle\exp\left(-\frac{m_c^2}{T^2}\right)\nonumber\\
 && -\frac{m_c\langle\bar{q}g_s\sigma Gq\rangle}{2T^2}\left(1-\frac{m_c^2}{2T^2}\right)\exp\left(-\frac{m_c^2}{T^2}\right) +\frac{1}{12}\langle \frac{\alpha_sGG}{\pi}\rangle\exp\left(-\frac{m_c^2}{T^2}\right)\nonumber\\
 &&-\frac{16\pi\alpha_s\langle\bar{q}q\rangle^2 }{27T^2}\left(1+\frac{m_c^2}{2T^2}
 -\frac{m_c^4}{12T^4}\right)\exp\left(-\frac{m_c^2}{T^2}\right)  \, ,
\end{eqnarray}
  \begin{eqnarray}
 f_{J/\psi}^2 M_{J/\psi}^2\exp\left(-\frac{M_{J/\psi}^2}{T^2}\right)&=& \frac{3}{4\pi^2} \int_{4m_c^2}^{s_0} ds \int_{y_i}^{y_f}dy \left\{y(1-y)\left(2s-\widetilde{m}_c^2\right) +3m_c^2\right\} \exp\left(-\frac{s}{T^2}\right)  \nonumber\\
 &&  +\frac{m_c^2}{24T^2}\langle \frac{\alpha_sGG}{\pi}\rangle\int_0^1 dy \left( 1-\frac{\widetilde{m}_c^2}{T^2}\right)\left\{ \frac{1-y}{y^2}+\frac{y}{(1-y)^2}\right\}\exp\left(-\frac{\widetilde{m}_c^2}{T^2}\right)\nonumber\\
  &&  -\frac{m_c^4}{24T^4}\langle \frac{\alpha_sGG}{\pi}\rangle\int_0^1 dy \left\{ \frac{1}{y^3}+\frac{1}{(1-y)^3}\right\}\exp\left(-\frac{\widetilde{m}_c^2}{T^2}\right)\nonumber\\
  &&  +\frac{m_c^2}{8T^2}\langle \frac{\alpha_sGG}{\pi}\rangle\int_0^1 dy \left\{ \frac{1}{y^2}+\frac{1}{(1-y)^2}\right\}\exp\left(-\frac{\widetilde{m}_c^2}{T^2}\right)\nonumber\\
  &&  -\frac{1}{12}\langle \frac{\alpha_sGG}{\pi}\rangle\int_0^1 dy \left( 1+\frac{\widetilde{m}_c^2}{2T^2}\right)\exp\left(-\frac{\widetilde{m}_c^2}{T^2}\right) \, ,
\end{eqnarray}
$y_{f}=\frac{1+\sqrt{1-4m_c^2/s}}{2}$, $y_{i}=\frac{1-\sqrt{1-4m_c^2/s}}{2}$, $ \widetilde{m}_c^2=\frac{m_c^2}{y(1-y)}$.
We derive Eqs.(26-27) with respect to $1/T^2$, then eliminate the decay constants $f_{D}$ and $f_{J/\psi}$ to obtain the QCD sum rules
for the masses  $M_{D}$ and $M_{J/\psi}$.
 We carry out the
operator product expansion to the vacuum condensates   up to dimension-6 in a consistent way and discard the perturbative corrections,
assume vacuum saturation for the  four-quark condensates \cite{Wang1301} and neglect the three gluon condensate so as to be consistent with the truncations in the operator product expansion in the QCD sum rules for the tetraquark  states.

The threshold parameters are chosen as  $s_0=6.2\,\rm{GeV}^2$ and $13\,\rm{GeV}^2$ for the $D$ and $J/\psi$ respectively according to
  the first radial excited states $D(2550)$ (or $D_J(2580)$) and $\psi^{\prime}(3686)$ \cite{PDG,D2550}.
   We usually take  the flavor $n_f=3$ and energy scale $\mu=\sqrt{m_D^2-m_c^2}\approx 1\,\rm{GeV}$ to study the $D$ meson. If larger energy scales are taken,
   for example, $\mu=(1.0-1.7)\,\rm{GeV}$, the experimental value $M_D=1.87\,\rm{GeV}$ can   be reproduced approximately with suitable Borel parameters $T^2$ in the region $ (1.6-2.3)\,\rm{GeV}^2$.
For the $J/\psi$, if the energy scales   $\mu=(1.1-1.6)\,\rm{GeV}$ are taken, the  experimental value $M_{J/\psi}=3.1\,\rm{GeV}$ can be reproduced
approximately with suitable Borel parameters $T^2$ in the region  $ (1.5-5.5)\,\rm{GeV}^2$. We have to bear in mind that such energy scales  and
truncations in the operator product expansion cannot reproduce the experimental values of the decay constants $f_{D}$ and $f_{J/\psi}$ \cite{Wang1301}.
If we only concern for  the masses, the acceptable   energy scales of the QCD sum rules for the hidden and open charmed mesons are about $\mu=(1.1-1.6)\,\rm{GeV}$.
For the tetraquark states, it is more reasonable to refer to the $\lambda_{X/Z}$ as the pole residues or couplings (not the decay constants).
We cannot obtain the true values of the pole residues $\lambda_{X/Z}$ by measuring the leptonic decays as in the cases of the $D_{s}(D)$ and $J/\psi (\Upsilon)$,
$D_{s}(D)\to \ell\nu$ and $J/\psi (\Upsilon)\to e^+e^-$, and have to calculate the  $\lambda_{X/Z}$ using  some theoretical methods, for example,
the lattice QCD. It is hard to obtain the true values. In this article, we focus on the masses to study the tetraquark states, and the predictions of the pole
residues maybe not as robust.

The threshold parameters of the axial-vector tetraquark states $X(3872)$ and $Z_c(3900)$ (or $Z_c(3885)$) are taken as $\sqrt{s_0}=(4.3-4.5)\,\rm{GeV}$  tentatively
 to avoid the contaminations of the higher  resonances and continuum states, here we have assumed that the energy gap between the ground states and
 the first radial excited states is about $(0.4-0.6)\,\rm{GeV}$, just like that of the conventional mesons.

\begin{figure}
\centering
\includegraphics[totalheight=6cm,width=7cm]{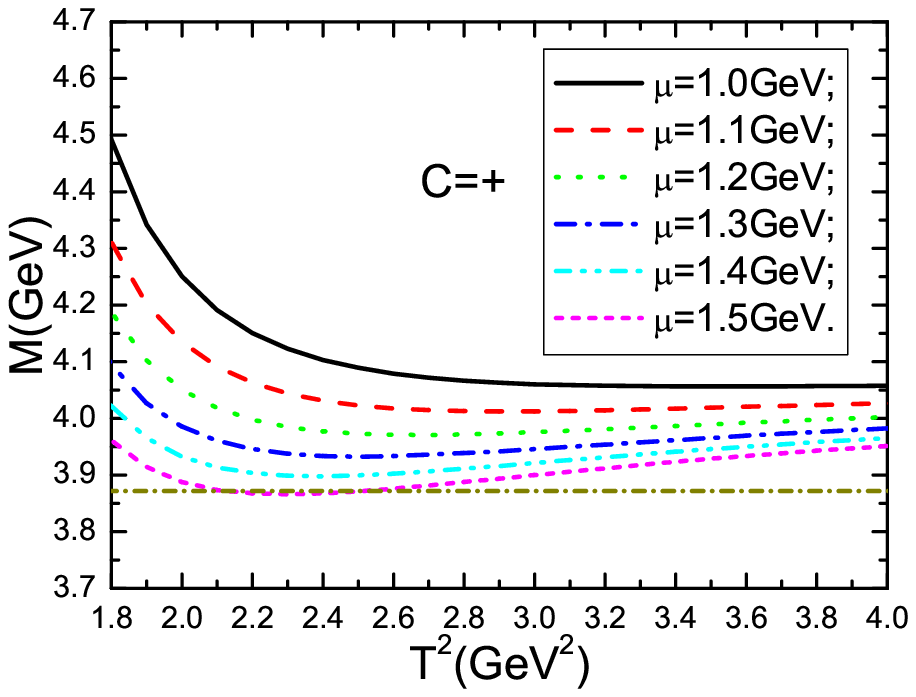}
\includegraphics[totalheight=6cm,width=7cm]{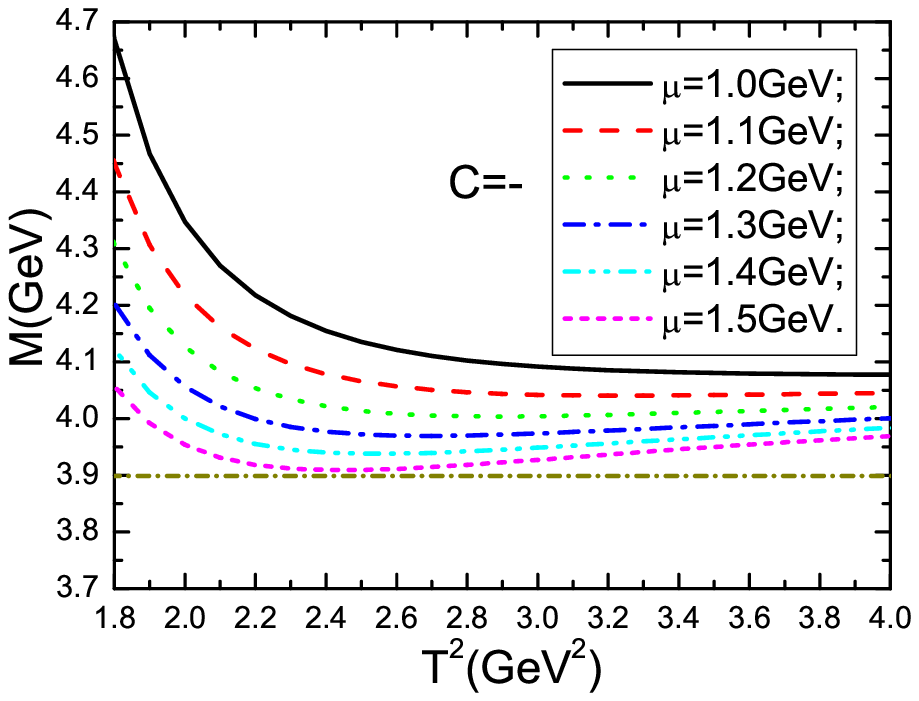}
  \caption{ The masses  with variations of the  Borel parameters $T^2$ and energy scales $\mu$, where the horizontal lines denote the experimental values. }
\end{figure}
In Fig.2,  the masses are plotted   with variations of the  Borel parameters $T^2$ and energy scales $\mu$ for the
threshold parameter $\sqrt{s_0}=4.4\,\rm{GeV}$. From the figure, we can see that the masses decrease monotonously
with increase of the energy scales. The energy scale $\mu=1.5\,\rm{GeV}$ is the lowest energy scale to reproduce the experimental data.

 \begin{figure}
\centering
\includegraphics[totalheight=6cm,width=7cm]{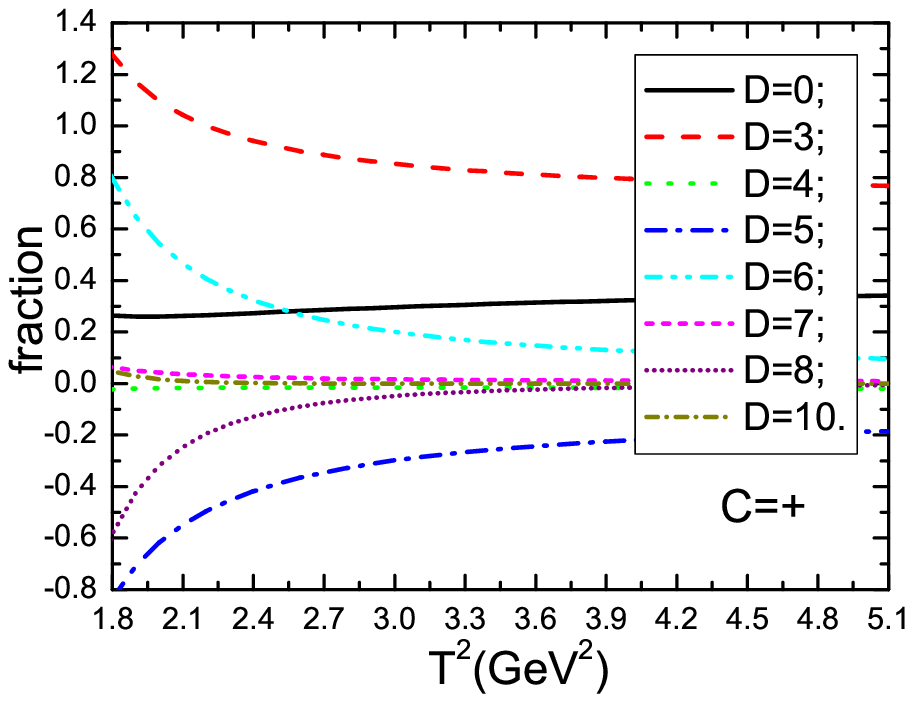}
\includegraphics[totalheight=6cm,width=7cm]{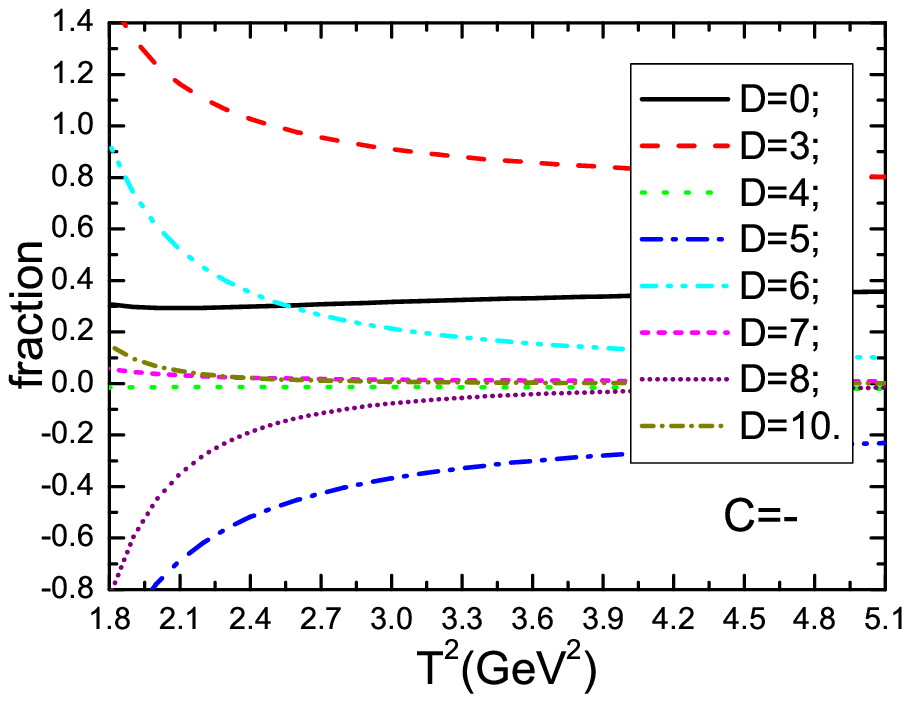}
  \caption{ The contributions of different terms in the operator product expansion  with variations of the
  Borel parameters $T^2$, where the $D$ denotes the dimensions of the vacuum condensates. }
\end{figure}
In Fig.3,  the contributions of different terms in the
operator product expansion are plotted with variations of the Borel parameters  $T^2$ for the threshold parameter $\sqrt{s_0}=4.4\,\rm{GeV}$ and energy scale $\mu=1.5\,\rm{GeV}$. From the figure, we can see that the
contributions  change quickly with variations of the Borel parameters
at the region $T^2< 2.1\,\rm{GeV}^2$, which does  not warrant
platforms for the masses. At the value $T^2=2.2\,\rm{GeV}^2$, the $D_0$, $D_3$, $D_4$, $D_5$, $D_6$, $D_7$, $D_8$, $D_{10}$  are
$0.266$,	$1.000$, $-0.017$, $-0.495$,	$0.406$,	$0.032$,	$-0.194$,	$0.006$ respectively for the $C=+$ tetraquark state;
$0.294$,	$1.106$,$-0.013$,	$-0.617$,	$0.450$,	$0.028$,	$-0.279$,	$0.036$ respectively for the $C=-$ tetraquark state,
where the $D_i$ with $i=0,\,3,\,4,\,5,\,6,\,7,\,8,\,10$ denote the contributions of the vacuum condensates of dimensions $D=i$, and the total contributions are normalized to be $1$.
Although the contributions of the condensates do not decrease monotonously with increase of dimensions, the $D_4$, $D_7$, $D_{10}$ play a less important role,
$D_3\gg |D_5|> D_6\gg D_8$,  the $D_6$, $D_8$, $D_{10}$ decrease monotonously and quickly  with increase of the Borel parameters.
The convergence of the operator product expansion does not mean that the perturbative terms make dominant contributions,
as the  continuum hadronic spectral densities  are approximated by
$\rho_{QCD}(s)\Theta(s-s_0)$ in the QCD sum rules for the tetraquark states, where
the $\rho_{QCD}(s)$ denotes the full QCD spectral densities; the contributions of the quark condensate $\langle\bar{q}q\rangle$ (of dimension-3) can be very large.
  In this article,  the
value $T^2\geq 2.2\,\rm{GeV}^2$ is taken tentatively, and the convergent
behavior in the operator product  expansion is very good.

\begin{figure}
\centering
\includegraphics[totalheight=6cm,width=7cm]{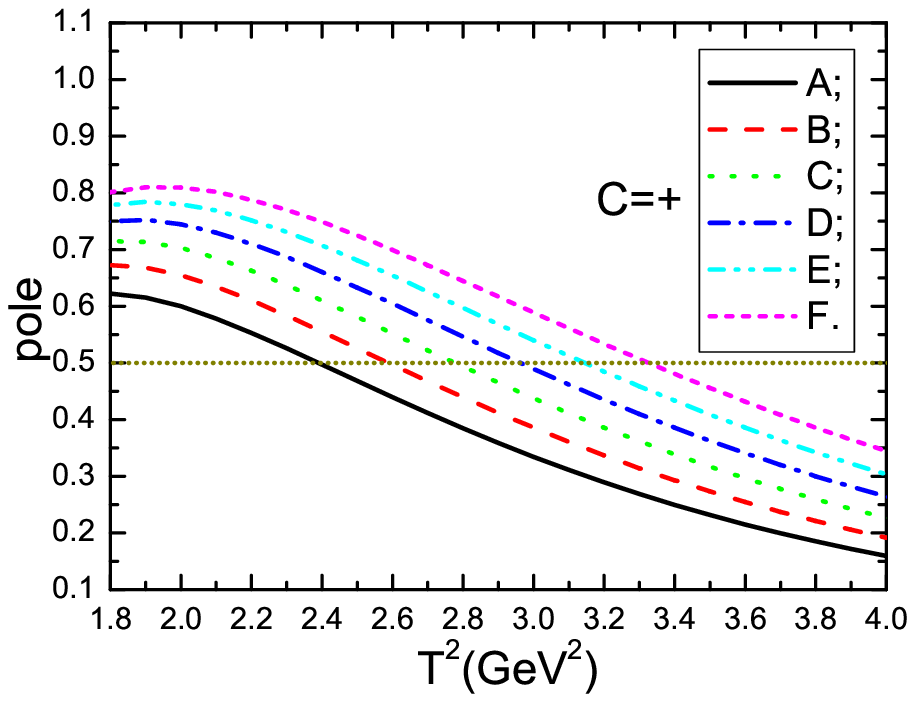}
\includegraphics[totalheight=6cm,width=7cm]{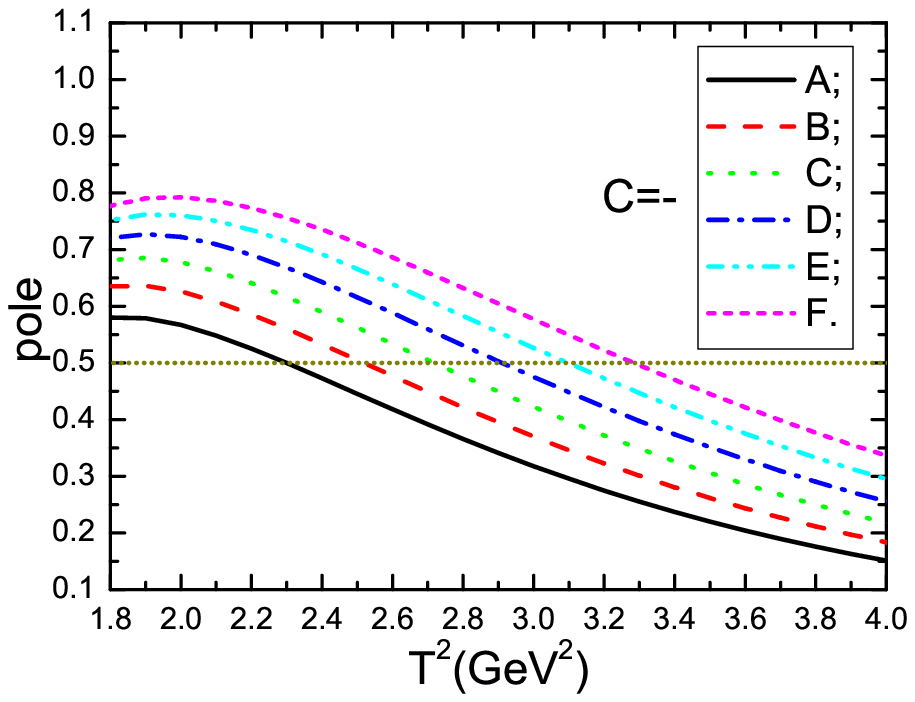}
  \caption{ The pole contributions  with variations of the  Borel parameters $T^2$ and threshold parameters $s_0$, where the $A$, $B$, $C$, $D$, $E$, $F$ denote the threshold parameters $\sqrt{s_0}=4.1$,  $4.2$, $4.3$, $4.4$, $4.5$, $4.6\,\rm{GeV}$, respectively. }
\end{figure}
In Fig.4,  the contributions of the pole terms are plotted with
variations of the threshold parameters $s_0$ and Borel parameters $T^2$ at the energy scale $\mu=1.5\,\rm{GeV}$. From the figure, we can
see that the values $\sqrt{s_0}\leq 4.2 \, \rm{GeV}$ are too small to
satisfy the pole dominance condition and result in reasonable Borel platforms. If we take the values
$\sqrt{s_0}=(4.3-4.5)\,\rm{GeV}$ and $T^2=(2.2-2.8)\,\rm{GeV}^2$, the pole
contributions are about $(49-75)\%$ and $(48-73)\%$ for the $C=+$ and $-$ tetraquark states respectively. The pole dominance condition is
well satisfied.

\begin{figure}
\centering
\includegraphics[totalheight=6cm,width=7cm]{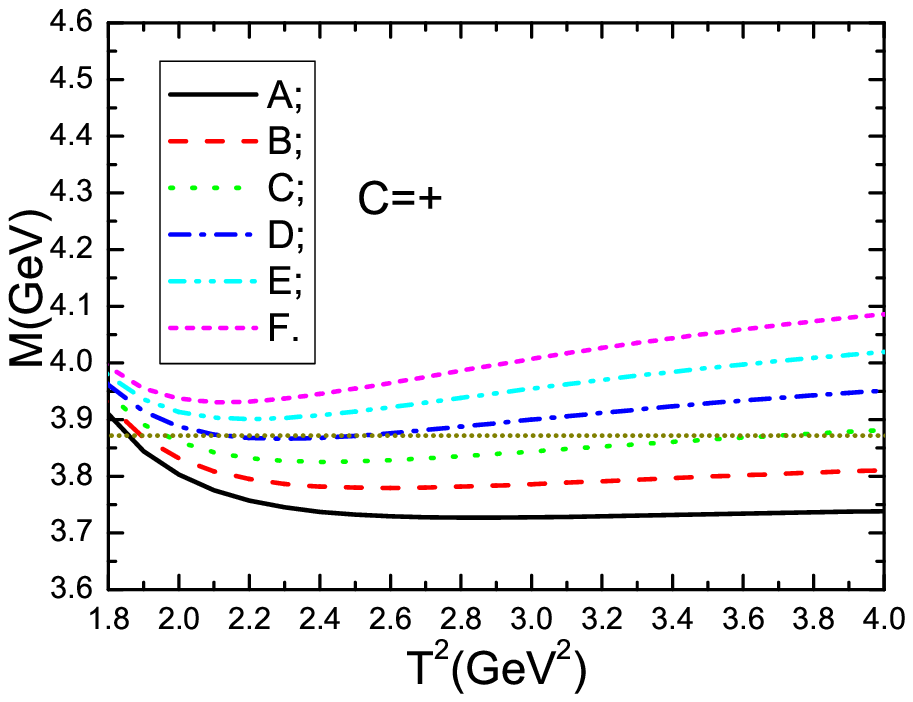}
\includegraphics[totalheight=6cm,width=7cm]{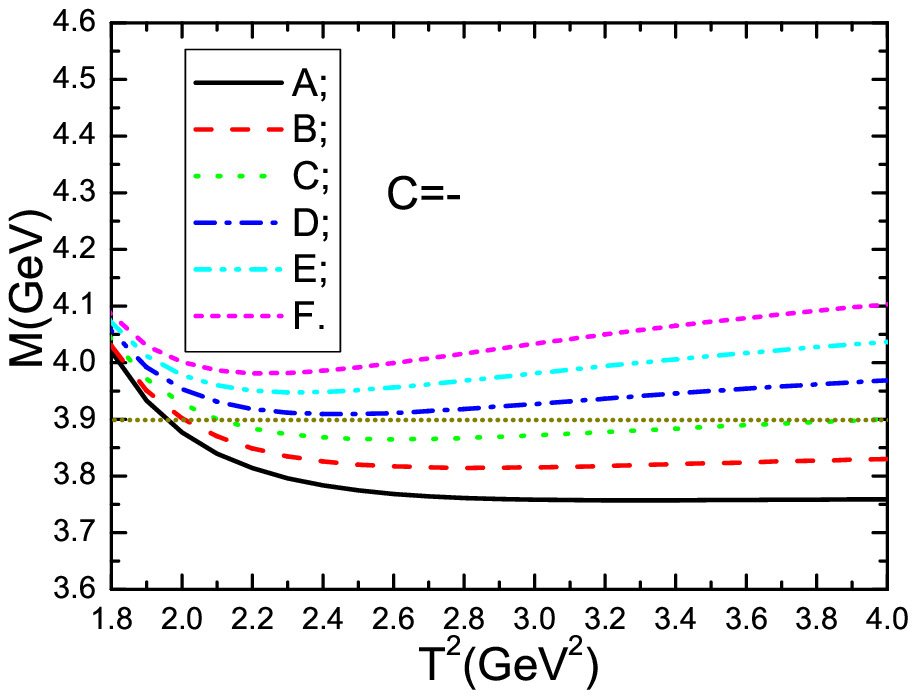}
  \caption{ The masses  with variations of the  Borel parameters $T^2$ and threshold parameters $s_0$, where the $A$, $B$, $C$, $D$, $E$, $F$ denote the threshold parameters $\sqrt{s_0}=4.1$,   $4.2$, $4.3$, $4.4$, $4.5$, $4.6\,\rm{GeV}$, respectively, and the horizontal lines denote the experimental values.  }
\end{figure}
In Fig.5,   the predicted masses are plotted with
variations of the threshold parameters $s_0$ and Borel parameters $T^2$ at the energy scale $\mu=1.5\,\rm{GeV}$. From the figure, we can
see that the value $\sqrt{s_0}= 4.4 \, \rm{GeV}$ is the optimal value to reproduce the experimental data. In this article,
   the parameters  $\sqrt{s_0}=(4.3-4.5)\,\rm{GeV}$, $T^2=(2.2-2.8)\,\rm{GeV}^2$ and $\mu=1.5\,\rm{GeV}$ are taken.

Taking into account all uncertainties of the input parameters,
finally we obtain the values of the masses and pole residues of
 the   $X(3872)$ and $Z_c(3900)$ (or $Z_c(3885)$), which are  shown explicitly in Figs.6-7,
\begin{eqnarray}
M_{X}&=&3.87^{+0.09}_{-0.09}\,\rm{GeV} \, ,  \nonumber\\
M_{Z}&=&3.91^{+0.11}_{-0.09}\,\rm{GeV} \, ,  \nonumber\\
\lambda_{X}&=&2.15^{+0.36}_{-0.27}\times 10^{-2}\,\rm{GeV}^5 \, , \nonumber\\
\lambda_{Z}&=&2.20^{+0.36}_{-0.29}\times 10^{-2}\,\rm{GeV}^5 \,   .
\end{eqnarray}
The uncertainties of the masses are very small, about $2.5\%$, as the uncertainties induced by the input parameters are canceled out to some extents between
 the numerators and denominators, see Eq.(24);
on the other hand, the uncertainties of the pole residues are much large, about $15\%$, as no cancelation  occurs among the induced uncertainties, see Eq.(19).
The prediction  $M_{X}=3.87^{+0.09}_{-0.09}\,\rm{GeV}$ is consistent with the experimental data $M_{X(3872)}=(3871.68\pm 0.17 )\,\rm{MeV}$ \cite{PDG},
and the prediction  $M_{Z}=3.91^{+0.11}_{-0.09}\,\rm{GeV}$ is also  consistent with the experimental data
$M_{Z_c(3900)}=(3899.0\pm 3.6\pm 4.9)\,\rm{ MeV}$ \cite{BES3900} and $M_{Z_c(3885)}=(3883.9 \pm 1.5 \pm 4.2)\,\rm{ MeV}$  \cite{BES-3885}
within uncertainties. The present predictions favor identifying the $X(3872)$ and $Z_c(3900)$ (or $Z_c(3885)$) as the $J^{PC}=1^{++}$ and $1^{+-}$
diquark-antidiquark type tetraquark states, respectively.
There is a small energy gap less than $40\,\rm{MeV}$ between the central values of the masses of the $C=+$ and $C=-$ axial-vector tetraquark states,
which is consistent with the value $10\,\rm{MeV}$ from the constituent diquark model \cite{Maiani-3872,Maiani1303}. The central values originate
from  the central values of all the input  parameters. We should bear in mind that the masses alone cannot qualify  the assignments ambiguously,
furthermore, the $M_{X}$ and $M_{Z}$ degenerate according to the uncertainties.

\begin{figure}
\centering
\includegraphics[totalheight=6cm,width=7cm]{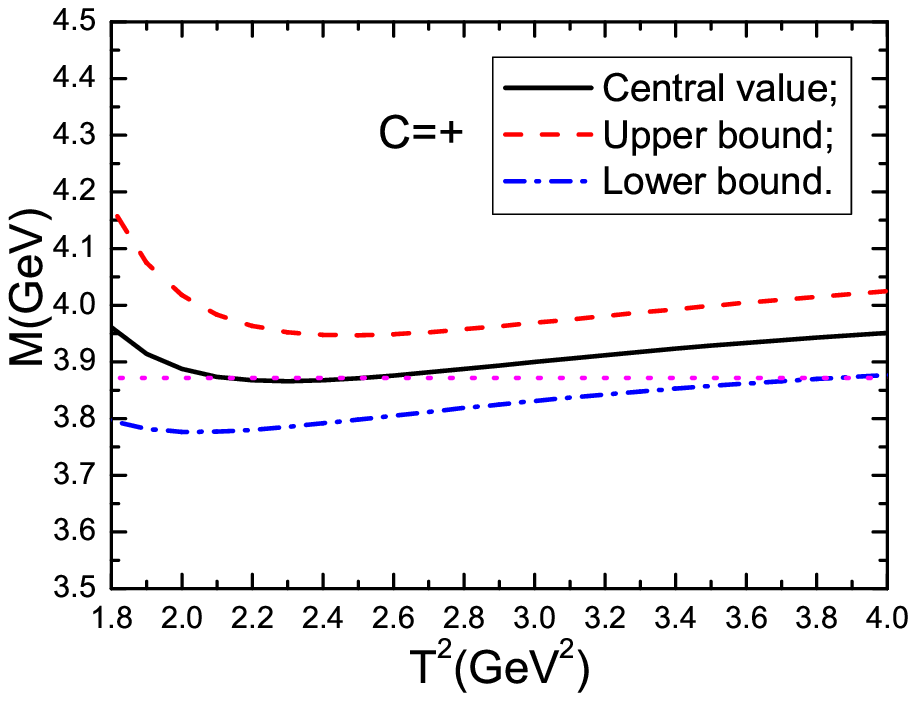}
\includegraphics[totalheight=6cm,width=7cm]{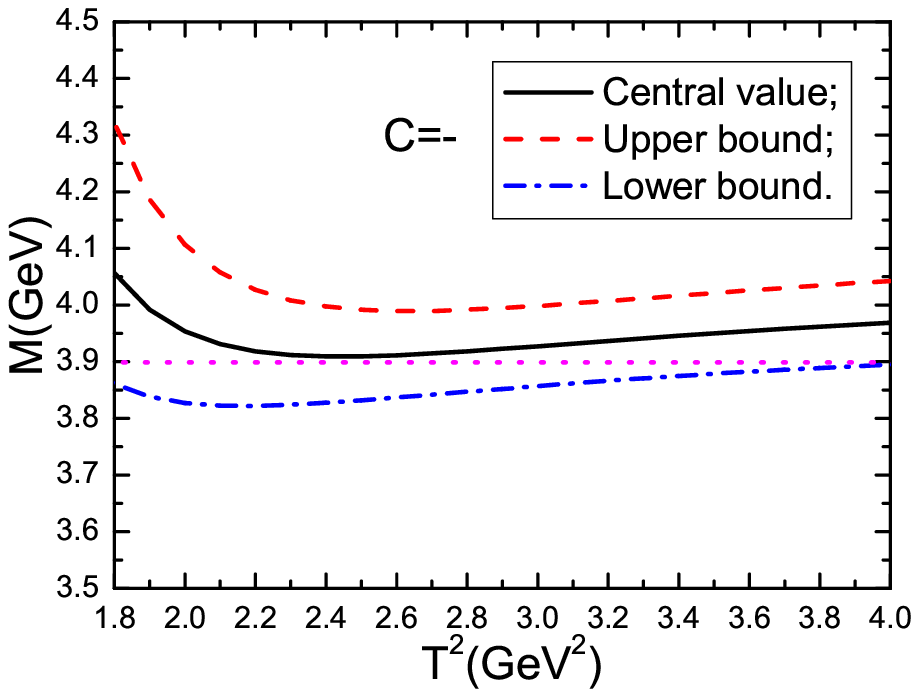}
  \caption{ The masses  with variations of the  Borel parameters $T^2$, where the horizontal lines denote  the experimental values.}
\end{figure}

\begin{figure}
\centering
\includegraphics[totalheight=6cm,width=7cm]{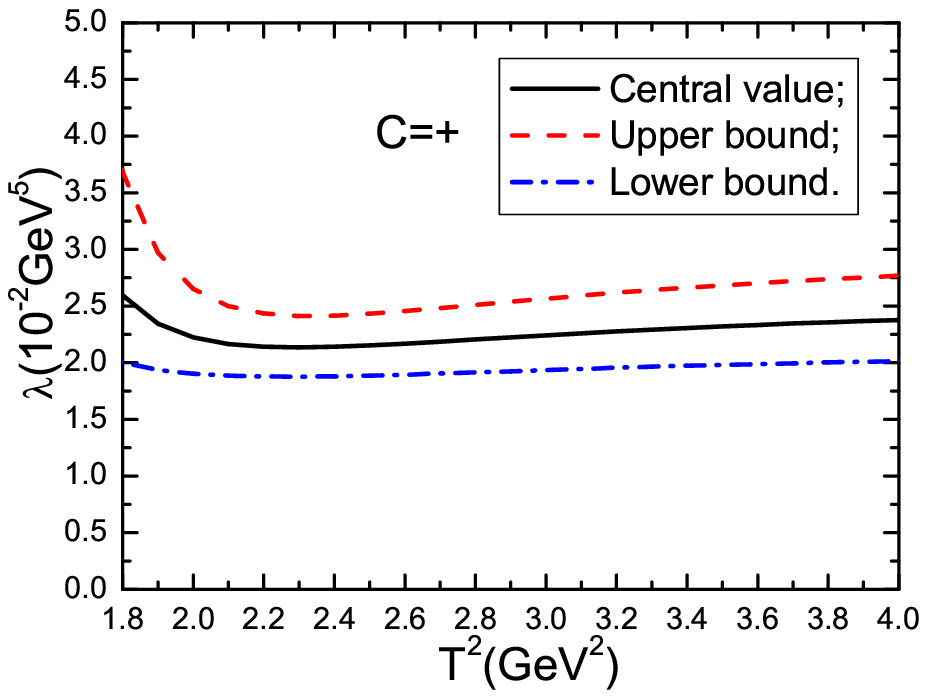}
\includegraphics[totalheight=6cm,width=7cm]{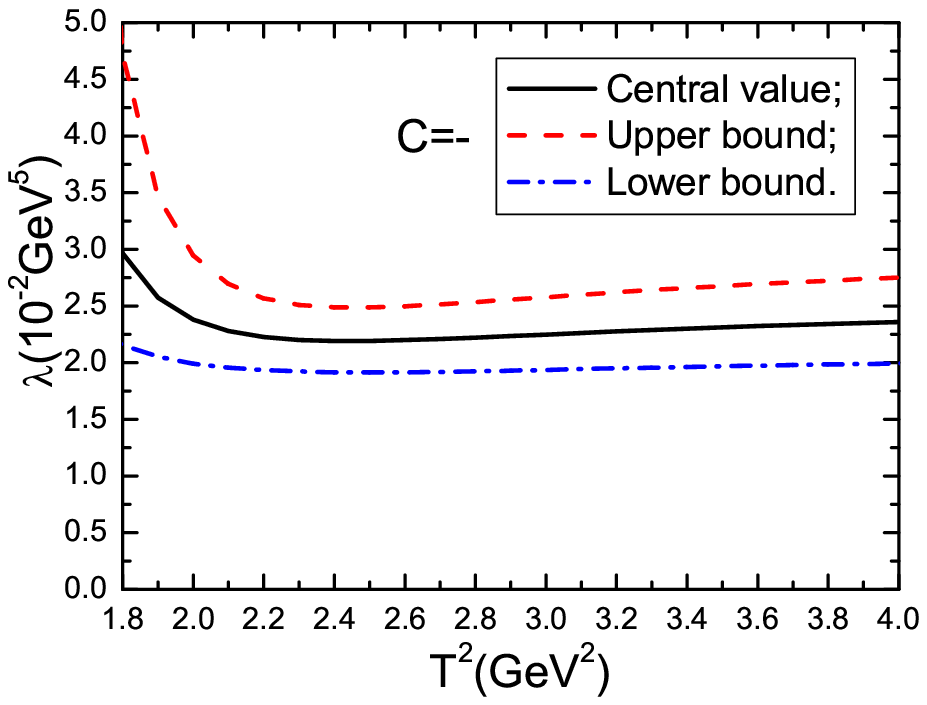}
  \caption{ The pole residues   with variations of the  Borel parameters $T^2$. }
\end{figure}

\section{Conclusion}
In this article, we  distinguish the charge conjunctions of the interpolating currents,  calculate the contributions of the vacuum condensates up to dimension-10 in a
consistent way in the operator product expansion and discard the perturbative corrections,  and take into account the higher dimensional vacuum condensates neglected in previous works, as they play an important role in determining the Borel windows. Then we study  the $J^{PC}=1^{+\pm}$ diquark-antidiquark type hidden charmed tetraquark states with the QCD sum rules, explore the energy scale dependence  in details  for the first time, and make reasonable predictions of the masses $M_{X}=3.87^{+0.09}_{-0.09}\,\rm{GeV}$, $M_{Z}=3.91^{+0.11}_{-0.09}\,\rm{GeV}$
 and pole residues $f_{X}=2.15^{+0.36}_{-0.27}\times 10^{-2}\,\rm{GeV}^5$, $f_{Z}=2.20^{+0.36}_{-0.29}\times 10^{-2}\,\rm{GeV}^5$. In calculations,
  we observe that  the tetraquark  masses decrease monotonously with increase of the energy scales, $\mu=1.5\,\rm{GeV}$ is the lowest energy scale to
  reproduce the experimental data.
 The  energy scale $\mu=1.5\,\rm{GeV}$ can also lead to reasonable  masses for the charmed mesons $D$ and $J/\psi$, and serves as an  acceptable
   energy  scale for the charmed mesons in the QCD sum rules.
The predictions support identifying    the $X(3872)$ and $Z_c(3900)$ (or $Z_c(3885)$) as the $1^{++}$ and $1^{+-}$ diquark-antidiquark type
tetraquark states, respectively. The  pole residues can be taken as
basic input parameters to study relevant processes of the $X(3872)$ and $Z_c(3900)$ (or $Z_c(3885)$) with the three-point QCD sum rules.

\section*{Acknowledgements}
This  work is supported by National Natural Science Foundation,
Grant Numbers 11375063, 11235005,  and the Fundamental Research Funds for the
Central Universities.

\end{document}